\documentclass[english]{sbrt}
\usepackage[english]{babel}
\usepackage[utf8]{inputenc}

\usepackage{amsmath}
\usepackage{amssymb}
\usepackage{xcolor}
\usepackage{graphicx}

\usepackage{subcaption}
\usepackage{caption}
\usepackage{pgfplots}
\usepgfplotslibrary{statistics} 
\pgfplotsset{compat=1.10}

\usepackage[%
maxcitenames=3,
mincitenames=1,
doi=false,
style=ieee,
backend=biber, 
]{biblatex}

\usepackage[nolist,printonlyused]{acronym}  

\begin{document}

\begin{acronym}[D-MIMO]
    
    \acro{3D}{three-dimensional}
    \acro{5G}{fifth generation}
    \acro{6G}{sixth generation}
    \acro{AP}{access point}    
    \acro{CDF}{cumulative distribution function}
    \acro{CIAIBS}{cumulative-individual-actual-INR-based selection}
    \acro{CIEIBS}{cumulative-individual-estimated-INR-based selection}
    \acro{CPU}{central processing unit}
    \acro{D-MIMO}{distributed multiple-input and multiple-output}
    \acro{DBZ}{distance-based zoning}
    \acro{FDD}{frequency division duplex}    
    \acro{FR3}{frequency range 3}
    \acro{FS}{fixed service}
    \acro{HPBW}{half power beamwidth}    
    \acro{IAIBS}{individual-actual-INR-based selection}
    \acro{IEIBS}{individual-estimated-INR-based selection}
    \acro{INR}{interference-to-noise ratio}
    \acro{ISD}{inter-site distance}
    \acro{ITU}{International Telecommunication Union}
    \acro{LOS}{line-of-sight}
    \acro{MMSE}{minimum mean square error}
    \acro{MNAU}{maximum number of admitted UEs}
    \acro{PN}{primary network}
    \acro{SE}{spectral efficiency}
    \acro{SISO}{single-input and single-output}
    \acro{SN}{secondary network}
    \acro{TDD}{time division duplex}
    \acro{UE}{user equipment}
    \acro{UMa}{urban macro}
    \acro{UPA}{uniform planar array}

\end{acronym}

\title{Interference Analysis in Spectrum Sharing Scenarios between Fixed Services and 6G Mobile Networks}

\author{Samuel S. Silva, Igor. M. Guerreiro, Yuri. C. B. Silva and Behrooz Makki
\thanks{Samuel S. Silva, Igor M. Guerreiro, and Yuri. C. B. Silva are with Wireless Telecom Research Group, Federal University of Ceará, Fortaleza, 
Brazil. E-mails: $\{$samuel.serejo, igor, yuri$\}$@gtel.ufc.br. Behrooz Makki is with Ericsson Research, Gothenburg, Sweden, E-mail: behrooz.makki@ericsson.com. 
This work was supported in part by Ericsson Research, Sweden, and Ericsson
Innovation Center, Brazil, Technical Cooperation Contracts UFC.53 and UFC.55, 
in part by CNPq, in part by CNPq/INCT-Signals Grant 406517/2022-3, 
and in part by CAPES - Finance Code 001.}%
}

\maketitle


\begin{abstract}

This study investigates the feasibility of coexistence between an incumbent \ac{FS} and a 6G network in the same location under spectrum sharing. For coexistence to be possible, a protection criterion must be met. Regulatory agencies define this criterion as the ratio of inter-network interference to noise at the \ac{FS} receiver. Simulation results indicate that this coexistence is not possible without the implementation of some interference mitigation strategy. To this end, a distance-based protection zoning and two user selection strategies were investigated from the fixed service perspective, assuming the 6G network is in uplink. The results indicate that meeting the protection criterion can be achieved when only few users in uplink are selected, while still providing optimistic capacity performance for such selected users in the shared band.

\end{abstract}

\section{Introduction}

The next generation of mobile communication systems~\cite{Jiang2021}, namely \ac{6G}, is expected to be capable of meeting a variety of requirements: higher data rates, lower latency in services, greater communication reliability, and better signal coverage. With the aim of meeting these requirements, several studies have investigated the use of new technologies and the implementation of new types of architectures. 

In this context, architectures in which antennas are distributed closer to users~\cite{Rendeiro2025, Demir_2021} are better suited to achieve such established goals than traditional cellular architecture. Also known as \ac{D-MIMO}, these networks are characterized by a large number of \acp{AP} across a given coverage area, which, in itself mitigates the problem of regions with poor signal coverage and allows the use of centralized processing to enhance decoding and preprocessing, as well as enabling joint and coherent transmissions, that increase signal levels and, consequently, improves quality of service.

Furthermore, \ac{6G} networks are also expected to occupy new portions of the frequency spectrum~\cite{Ericsson2024}, acquiring wider bandwidth to meet the capacity demands and improve service quality. 
Unlike what occurred with the \ac{5G}, attention is now turning to the portion of the spectrum within 7 to 15 GHz: a frequency band known as \ac{FR3}. This band corresponds to centimeter-wave frequencies. The factors driving interest in utilizing this band are greater availability compared to the sub-6 GHz band and the better propagation conditions when compared to the millimeter-wave bands adopted in \ac{5G}. 

However, there is an obstacle to the practical use of \ac{FR3} bands by upcoming wireless systems: this portion of the spectrum is already heavily allocated to various incumbent services. These incumbent services include satellite communications, fixed links, military applications, radar services, and many others~\cite{Bazzi2026}. Access to \ac{FR3} bands for mobile networks is then granted through regulatory measures implemented by international organizations, regional initiatives, or local regulatory agencies, which establish some protection criteria to ensure that the incumbent services are not affected by a sudden increase in external interference levels.

For fixed services in \ac{FR3}, without more detailed specifications, the \ac{ITU} protection criterion required for 
long-term interference must be significantly below the noise floor most of the time~\cite{ITUR_F758_8}.

Therefore, several studies have investigated different network aspects~\cite{Tercero2016,Nokia2024} and 
transmission strategies~\cite{Shaik2024,Paula2025} in coexistence scenarios. Among these strategies are beamforming techniques, to focus beams towards users and avoid causing interference in the direction of the incumbent network; power control, to increase the system’s energy efficiency and reduce the level of interference caused. Each strategy is designed for different levels of cooperation or mutual understanding between the networks: for some applications, the locations of the receiver and fixed transmitter are known to the regulator, as are power parameters and others. 
Furthermore, most of these studies assume a low—or nonexistent—level of cooperation and also that the networks are physically far apart.

With this in mind, this study conducts an analysis of the impact of interference arising from sharing a portion of the spectrum between an incumbent fixed service and a \ac{D-MIMO} mobile network in uplink. We assume that both networks reside in the same area in close proximity to one another. No communication or coordination between the two networks is assumed. Since the fixed service holds the right to use the spectrum, the \ac{D-MIMO} mobile network must adapt to the environment, i.e., it must avoid causing significant interference to the incumbent service. 

This work then examines how simple techniques, such as protection zoning and user selection, under different criteria, can help mitigate inter-network interference levels and meet established protection criteria. The impact of these strategies on relevant key performance indicators is also investigated.

\section{System Model}
\label{sec:sys_model}

\begin{figure}[!t]
    \centering
    \includegraphics[scale=0.5]{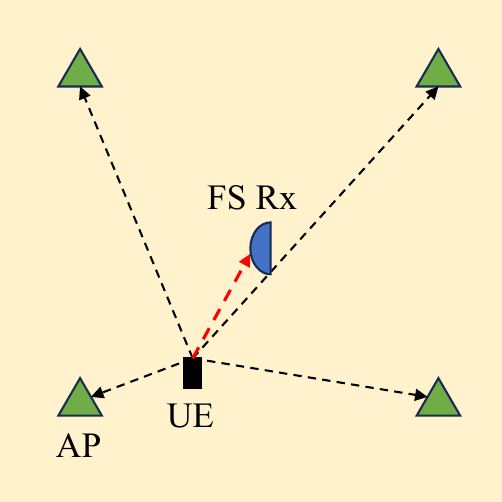}

    \caption{Illustrative coexistence scenario with an incumbent service, i.e., the \ac{FS} link and a \acs{D-MIMO} \ac{SN} with \acsp{AP} 
    serving a \acs{UE} in uplink.}
    \label{fig:scenario}
\end{figure}

We investigate a scenario in which an incumbent \ac{FS} -- the \ac{PN} -- and a \ac{D-MIMO} mobile network -- the \ac{SN} -- coexist in the same location and share the same spectrum band, as illustrated in Fig.~\ref{fig:scenario}. The \ac{PN} and the \ac{SN} operate in \ac{FDD} and \ac{TDD} modes, respectively.

The \ac{PN} consists of a single fixed link, with the two ends separated by a long distance. Both the transmitter and receiver are equipped with highly directional antennas, and their link features a strong \ac{LOS}. The \ac{FS} receiver is located at the center of the coverage area, and is thus surrounded by elements of the other network, making it more susceptible to inter-network interference.

The \ac{SN} consists of $K$ single-antenna \ac{UE} devices and $L$ multi-antenna \acp{AP}. The \acp{UE} are randomly positioned across the coverage area, while the \acp{AP} are disposed in a regular grid with an \ac{ISD} of 350 meters. The \acp{UE} are equipped with a single omni antenna and each \ac{AP} is equipped with 3 \ac{UPA} panels, each consisting of $N$ antennas and with their respective antenna array boresights separated by 120 degrees. In addition, all \acp{AP} are connected to a single \ac{CPU} through error-free fronthaul links for centralized signal processing purposes. 

Given the general scenario described above, this work focuses on the impact of \ac{SN} uplink transmissions on the \ac{FS} receiver. In addition, to model more realistic propagation conditions and ensure the directional characteristics inherent in fixed links, this study considers the radiation patterns in~\cite{3gpp38901} and~\cite{itur_f699} for the \ac{SN} \acp{AP} and for the \ac{FS} link, respectively.

\subsection{Propagation Model}

The large-scale channel coefficients can be described as:

\begin{equation}
    \beta_{k,l} = 10^{\frac{\mathrm{PL}_{k,l} + \mathrm{SH}_{k,l}}{10}},
\end{equation}

\noindent where $\mathrm{PL}_{k,l}$ is the path-loss and $\mathrm{SH}_{k,l}$ is the spatial correlated shadowing, 
both in dB, with standard deviation $\sigma_{\text{SH}}$, which varies depending on the \ac{LOS} condition of the link. 

The channel between an \ac{SN} \ac{UE} and any receiver, which can be either an \ac{SN} \ac{AP} or an \ac{FS} receiver, is modeled following a Rician distribution. All links are subject to the possibility of having a direct path between the transmitter and the receiver, which depends on their positions and heights and directly affects propagation conditions. The path-loss calculation, shadowing effect generation and the Rician K-factor are entirely dependent on whether or not the channel is \ac{LOS}. 

If the link has \ac{LOS}, $\kappa \sim \mathcal{N}(5, 9)$ dB; otherwise, $K$ is equal to zero. The expressions for the probability of \ac{LOS}, path loss, and the variance of the shadowing effect are all consistent with the expressions available in  \cite{3gpp38901} for \ac{UMa} scenarios.

\subsection{\ac{UE}-\ac{AP} Channel Response}

For the \ac{UE}-\ac{AP} link in the \ac{SN}, the channel response $\mathbf{h}_{k,l} \in \mathbb{C}^{N}$ between 
the $k$-th \ac{UE} antenna and the $l$-th \ac{AP} receive antenna array can be expressed as follows:

\begin{equation}
    \mathbf{h}_{k,l} = \left[ 
    \sqrt{ \frac{\kappa_{k,l}} {\kappa_{k,l} + 1}} \mathbf{a}_{k,l} 
    + \sqrt{ \frac{1}{\kappa_{k,l} + 1}} \mathbf{h}^{\omega}_{k,l}
    \right] \sqrt{\beta_{k,l} G_{k,l}}\,,
\label{eq:ch_ue_ap}
\end{equation}

\noindent in which $\kappa_{k,l}$ denotes the Rician K-factor; $\mathbf{h}_{k,l}^{\omega}$ is the diffuse 
component that follows a complex Gaussian distribution, i.e., $\mathbf{h}_{k,l}^{\omega} \sim \mathcal{CN}\left(\mathbf{0}, \mathbf{R}_{k,l} \right)$, with $\mathbf{R}_{k,l} \in \mathbb{C}^{N \times N}$ being the spatial correlation matrix; $\mathbf{a}_{k,l} \in \mathbb{C}^{N}$ is the array steering vector~\cite{balanis2015antenna}; $\beta_{k,l}$ is the large scale fading coefficient and $G_{k,l}$ is the radiation power pattern.

The gain of a single antenna element in a \ac{SN} \ac{AP} antenna array can be calculated, for both the vertical and horizontal axes, using the model described in~\cite{3gpp38901}. The gain $G \left( \theta\right)$ of this respective element can be expressed as:

\begin{equation}
    G \left( \theta\right) = - \text{min} \left[ 
    12 \left( \frac{\theta}{\theta_{3 \text{dB}}}\right)^2, A_m
    \right]\,,
    \label{eq:rpp}
\end{equation}

\noindent in which $\theta$ is the angle of incidence of the incoming wave, relative to the direction in which 
the antenna is pointing along that axis, $\theta_{3\text{dB}}$ is the \ac{HPBW} and $A_m = 20\ \text{dB}$ is the maximum attenuation of the antenna gain. For the vertical axis, the relative angle of incidence $\theta$ is contained within $\theta \in [0, \frac{\pi}{2}]$, while for the horizontal axis, the relative angle of incidence is contained within $\phi \in [0, \pi]$. Consequently, the \ac{3D} radiation power pattern $G_{k,l}$ of the $(k,l)$-th \ac{UE}-\ac{AP} link is given by:

\begin{equation}
    G_{k,l} = - \text{min} \left[ - \left(G_V\left(\theta_{k,l}\right) + G_H\left(\phi_{k,l}\right)\right), A_m \right]\,,
\end{equation}
where $G_V\left(\theta_{k,l}\right)$ and $G_H\left(\phi_{k,l}\right)$ are the vertical and the horizontal gains, 
as in~\eqref{eq:rpp}, evaluated at the two angles of incidence, $\theta_{k,l}$ and $\phi_{k,l}$ of the 
corresponding $(k,l)$-th \ac{UE}-\ac{AP} link.

\subsection{\ac{UE}-\ac{FS} Receiver Channel Response}

Regarding the \ac{UE}-\ac{FS} receiver links, the \ac{SISO} channel response $h_{k,\text{fs}}\in\mathbb{C}$ between 
the $k$-th \ac{UE} and the \ac{FS} receiver is similar to~\eqref{eq:ch_ue_ap} and can be expressed as:

\begin{equation}
    h_{k,\text{fs}} = \left[\sqrt{\frac{\kappa_{k,\text{fs}}} {\kappa_{k,\text{fs}} + 1}} + \sqrt{\frac{1} {\kappa_{k,\text{fs}} + 1}}h^{\omega}_{k,\text{fs}}\right]\sqrt{\beta_{k,\text{fs}} G_{k,\text{fs}}}\,,
\label{eq:ch_ue_fs}
\end{equation}
in which $K_{k,\text{fs}}$ denotes the Rician K-factor; $h_{k,\text{fs}}^{\omega} \sim \mathcal{CN}\left(0, 1 \right)$ is the random component; $\beta_{k,\text{fs}}$ is the large scale fading coefficient and $G_{k,\text{fs}}$ is the radiation power pattern.

This work follows~\cite{itur_f699} to model the radiation power pattern of the \ac{FS} antenna. To use the available expressions, it is necessary to know at least one of the parameters: the antenna’s maximum gain or the ratio of diameter to wavelength $D/\lambda$ at the operating frequency. Knowing the value of one allows you to determine the value of the other, and vice versa. We define the maximum gain of the \ac{FS} antenna as 47.4 dBi and use the following approximation to determine the ratio $D/\lambda$:

\begin{equation}
    G_{\text{max}} \approx 20 \log \frac{D}{\lambda} + 7.7\,.
\end{equation}

Using this expression, we were able to obtain an approximate value of $D/\lambda \approx 96.6$ for the required ratio. Following this, a radiation pattern was implemented that is compatible with antennas having a ratio of 100 or 
less and operating frequencies between 1 and 100 GHz, as described below: 

\begin{equation}
    G_{\text{fs}} \left( \theta \right) = 
    \begin{cases}
        G_\text{max} - 2.5 \times 10^{-3} \left( \frac{D}{\lambda} \theta\right)^2 & \text{if } 0 < \theta < \theta_m \,, \\

        G_1 & \text{if } \theta_\text{m} \leq \theta < 100 \frac{\lambda}{D} \,, \\ 

        52 - 10 \log\frac{D}{\lambda} - 25 \log\theta & \text{if } 100\frac{\lambda}{D} \leq \theta < 48 \,, \\

        10 - 10\log\frac{D}{\lambda} &\text{if } 48 \leq\theta \leq 180 \\ 
    \end{cases} \,,
    \label{ITU_rad}
\end{equation}

\noindent where $\theta$ is the angle of incidence of the signal relative to the angular direction toward which the antenna is pointing, $G_1$ is the gain of the first side lobe, with a value defined by $G_1 = 2 + 15\ \log\ \frac{D}{\lambda}$, and $\theta_m$ is an angular limit. The value of $\theta_\text{m}$ is given by the following expression:

\begin{equation}
    \theta_m = \frac{20 \lambda}{D} \sqrt{G_\text{max} - G_1}\,.
\end{equation}

The expression described in~\eqref{ITU_rad} refers to the gain of the dish antenna along only one of the axes: azimuthal or vertical. Finally, the \ac{3D} antenna pattern for the link between the $k$-th \ac{UE} and \ac{FS} receiver is given by:

\begin{equation}
    G_{k,\text{fs}} = G_{\text{fs},V}(\theta_{k,\text{fs}}) + G_{\text{fs},H}(\phi_{k,\text{fs}})\,.
    \label{eq:rpp_fs}
\end{equation}

\subsection{Interference Measurement}

In this study, we will limit ourselves to evaluating the inter-network interference perceived by the \ac{FS} receiver. The interference signal power received by the \ac{FS} receiver's antenna can be expressed by the following equation: 

\begin{equation}
    I =  \left| \sum^{K}_{i = 1} h_{i,\text{fs}} x _i \right|^{2}\,,
    \label{eq:interf}
\end{equation}

\noindent in which $h_{i,\text{fs}}$ is the channel response defined in~\eqref{eq:ch_ue_fs}, $x_i$ is the transmitted signal from the $i$-th \ac{UE} so that $\mathrm{E}[|x_i|^2]=\rho_i$, with $\rho_i$ standing for its maximum transmission power.

Based on~\eqref{eq:interf}, the \ac{INR} experienced by the \ac{FS} receiver can be calculated as follows:
\begin{equation}
    \Gamma = \frac{I}{\sigma_\text{fs}^2}\,,
    \label{eq:inr}
\end{equation}
in which $\sigma_\text{fs}^2$ is the noise variance at the \ac{FS} receiver. Such a metric is used to assess the impact of inter-network interference on the \ac{FS} link.

\section{Interference Mitigation strategies}

To enable the coexistence between the two networks without significantly disrupting the incumbent service, 
international telecommunications organizations, e.g., \ac{ITU}, establish protection criteria for incumbent service providers, which should be used in studies. According to~\ac{ITU}, the long-term protection criterion for fixed services is an \ac{INR} of -10 dB or less for 80\% of the time. 

Therefore, in the following, three different interference mitigation strategies at the \ac{SN} are evaluated 
to avoid violation of the \ac{INR} threshold.

\subsection{Distance-based Protection Zoning}

The \ac{DBZ} strategy involves defining circular protection zones with radius $r$ around the \ac{FS} receiver. 
Any user located within these zones is not allowed to transmit on the same frequency band as the \ac{FS} receiver. 
This strategy relies only on the knowledge of the positions of the \ac{FS} receiver and the \acp{UE} by the \ac{SN}.

\subsection{Individual-\ac{INR}-based Selection Strategy}

The \ac{IAIBS} strategy relies on the knowledge of per-\ac{UE} \acp{INR}, defined as:
\begin{equation}
    \Gamma_k = \frac{|h_{k,\text{fs}} x_k|^2}{\sigma_\text{fs}^2}\,.
    \label{eq:inr_indiv}
\end{equation}
From~\eqref{eq:inr_indiv}, the \ac{SN} only selects \acp{UE} whose \acp{INR} are below a predetermined threshold to transmit on the same frequency as the \ac{FS} receiver.

In practice, it is not reasonable to assume that $\Gamma_k$ are known by the \ac{SN}. Therefore, 
an alternative to $\Gamma_k$ is its long-term version, \ac{IEIBS} strategy, that depends only on large-scale propagation effects, herein defined as:
\begin{equation}
    \bar{\Gamma}_k = \frac{\rho_k\beta_{k, \text{fs}} G_{k, \text{fs}} }{\sigma_\text{fs}^2}\,,
    \label{eq:inr_indiv_est}
\end{equation}
in which $\rho_k$ is the transmitted power of the $k$-th \ac{UE}. Assuming that the \ac{SN} has knowledge of 
the positions of \acp{UE} and the \ac{FS} receiver, it can obtain/estimate $\beta_{k, \text{fs}}$ and $G_{k, \text{fs}}$.

\subsection{Cumulative-Individual-\acp{INR}-based Selection Strategies}

This third strategy involves sorting users in ascending order by their individual \acp{INR}, either $\Gamma_k$, referred to as \ac{CIAIBS}, or $\bar{\Gamma}_k$, then \ac{CIEIBS}. Now let $\mathcal{U}$ be a subset of selected users. Assuming $\mathcal{U}$ initially empty, the \ac{SN} keeps adding \acp{UE} to $\mathcal{U}$, following the users' ascending order, so that the constraint below is not violated:
\begin{equation}
    \sum_{i \in \mathcal{U}} \Gamma_i \leq \gamma\,,
\end{equation}
in which $\gamma$ is a predetermined threshold. At the end, \acp{UE} in the resulting $\mathcal{U}$ are selected to 
transmit on the same frequency as the \ac{FS} receiver.

\section{Simulation Results}

The coexistence scenario studied in this work was simulated according to the system model in Section~\ref{sec:sys_model} and using 200 Monte Carlo simulations. The results were obtained by analyzing i) the \ac{INR} caused to the \ac{FS} receiver from the \ac{SN}, ii) the sum of \ac{SN} \ac{SE}, in terms of empirical \acp{CDF}, and iii) the number of \ac{SN} users that were admitted to operate on the shared frequency band.

The uplink sum \ac{SE} at the \ac{SN} is calculated based on~\cite{Demir_2021}, assuming joint processing, 
pilot-based channel estimation with no pilot contamination by adopting the \ac{MMSE} estimator, and signal decoding at the \ac{CPU} with a centralized \ac{MMSE} combiner. For simplicity, inter-network interference terms in the uplink sum \ac{SE} are neglected as this work assumes that the \ac{FS} transmitter is far from the \acp{UE} and the \ac{FS} link is extremely directional, according to~\eqref{eq:rpp_fs}. 

The simulated scenario comprises 8 \ac{SN} \acp{UE}, uniformly dropped on the coverage area, and served by 36 \acp{AP} placed in a regular grid. Each \ac{UE} is equipped with a single omni-directional antenna, and each \ac{AP} is equipped with a \ac{UPA} of $N = 32$ elements, with $N_x = 8$ and $N_y = 4$. The \ac{AP} arrays have a downtilt of 3 degrees. Table \ref{tab:tabela_de_parametros} shows the remaining parameters used in this work for the simulation of the system and propagation model. 

\begin{table}[t!]
\centering
\caption{Simulation parameters.}

\begin{tabular}{ll}
\hline
\textbf{Parameter} & \textbf{Value} \\ 
\hline
    Operating frequency &  $f_c = 7.5\ \text{GHz}$ \\
    Communication bandwidth & $B = 100\ \text{MHz}$ \\
    \ac{UE}/\acp{AP}/\ac{FS} heights & 1.5, 18, 50 m \\
    Number of pilots (ch. estimation) & $\tau_p = 10$ \\
    UL power per \ac{UE} & $\rho_k = 100$ mW \\
    Noise power density & -174 dBm/Hz \\
    Noise figure & 7 dB \\
    Shadowing variance (LOS, NLOS) & 4, 7.2 \\
    \ac{FS} antenna gain & 47.4 dBi \\
    \ac{SN} \acp{AP} antenna gain & 8 dBi \\    
\hline
\end{tabular}
\label{tab:tabela_de_parametros}
\end{table}

\definecolor{hungup}{HTML}{FFB6C1}    
\definecolor{crimsonred}{HTML}{DC143C} 
\definecolor{sorrymagenta}{HTML}{8B008B} 
\definecolor{slateblue}{HTML}{6A5ACD}   

\begin{figure}[!t]
    \centering
    \begin{subfigure}[b]{\columnwidth}
        \includegraphics{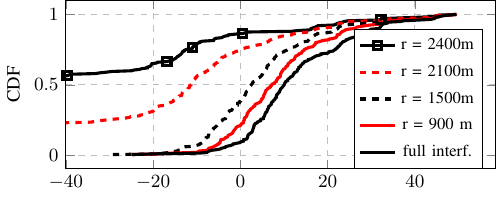}
    \caption{\ac{DBZ}}
    \label{fig:INR_for_zoning}
    \end{subfigure}

    \begin{subfigure}[b]{\columnwidth}
        \includegraphics{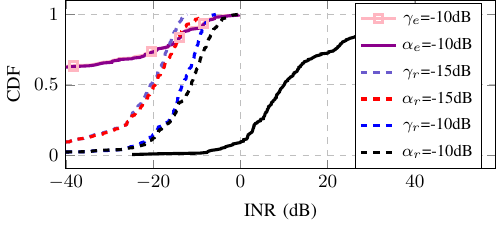}
    \caption{\ac{IEIBS}, \ac{IAIBS}, \ac{CIEIBS} and \ac{CIAIBS}}
    \label{fig:INR_for_selection}
    \end{subfigure}

    \caption{\acp{CDF} of \ac{INR} levels caused by \acp{UE} for the \ac{FS} receiver.}
    \label{fig:INR_levels_for_all_methods}
\end{figure}

\begin{figure}[!t]
    \centering
\begin{subfigure}[b]{\columnwidth}
    \centering
    \includegraphics{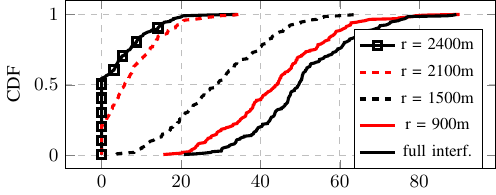}
    \caption{\ac{DBZ}}
    \label{fig:sum_of_SE_for_zoning}
\end{subfigure}

\begin{subfigure}[b]{\columnwidth}
    \centering
    \includegraphics{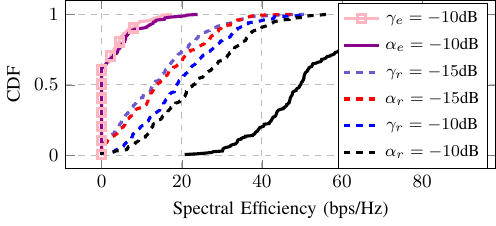}
    \caption{\ac{IEIBS}, \ac{IAIBS}, \ac{CIEIBS} and \ac{CIAIBS}}
    \label{fig:sum_of_SE_for_selection}
\end{subfigure}

\caption{\acp{CDF} of the sums of SN \acs{SE} for the different strategies.}
\label{fig:sum_of_SEs_for_all_methods}
\end{figure}

\begin{figure}[!t]
    \centering

    \begin{subfigure}[b]{\columnwidth}
        \centering
        \includegraphics{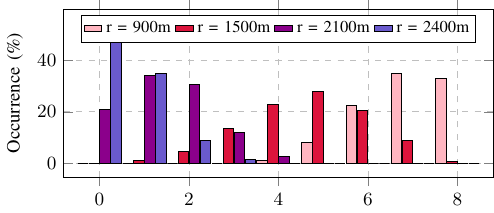}
    \caption{\ac{DBZ}.}
    \label{fig:Num_admitted_UEs_for_zoning}
    \end{subfigure}

    \begin{subfigure}[b]{\columnwidth}
        \centering
        \includegraphics{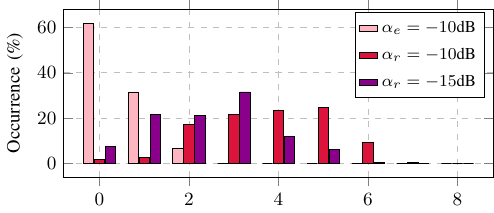}
    \caption{\ac{IEIBS} and \ac{IAIBS}}
    \label{fig:Num_admitted_UEs_for_selection}
    \end{subfigure}
    
    \begin{subfigure}[b]{\columnwidth}
        \centering
        \includegraphics{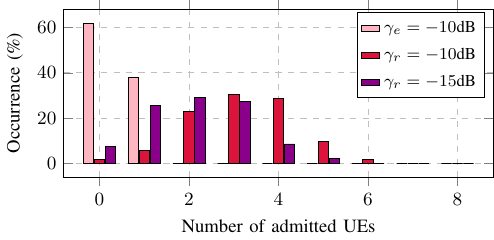}
    \caption{ \ac{CIEIBS} and \ac{CIAIBS} }
    \label{fig:Num_admitted_UEs_for_aggregate}
    \end{subfigure}

\caption{Histogram of the number of users admitted by each strategy.}
\label{fig:Num_admitted_UEs_for_all_methods}
\end{figure}

Figures \ref{fig:INR_levels_for_all_methods} and \ref{fig:sum_of_SEs_for_all_methods} show the resulting \ac{INR} levels and sums of \acp{SE} of \ac{SN} \acp{UE}, respectively, while Figure~\ref{fig:Num_admitted_UEs_for_all_methods} shows the number of \acp{UE} allocated on the shared band for the various interference mitigation strategies. From these results, it is clear that all techniques resulted in a reduction in the caused \ac{INR} levels, as well as in the sum of \acp{SE} in the shared band and the number of \acp{UE} allocated to it.

In Figure~\ref{fig:INR_levels_for_all_methods}, at the 80th percentile, for the baseline scenario the obtained \ac{INR} level was 23.325 dB. For \ac{DBZ} method none of the tested radii for the \ac{DBZ} technique achieve the required \ac{INR} reduction. For the \ac{DBZ} method with radius of 900, 1,500, 2,100 and 2,400 meters the obtained \ac{INR} levels were 17.29, 14.21, 5.79 and -9.3145 dB. The \ac{IEIBS} method with $\alpha_e = -10$ dB reached an \ac{INR} value of -15.17 dB. The \ac{IAIBS} method with $\alpha_r$ equal to -10 and -15 dB achieve \ac{INR} levels of -7.65 and -14.56 dB, respectively. The \ac{CIEIBS} technique with $\gamma_e= -10$ dB reached an \ac{INR} level of -15.168 dB. The \ac{CIAIBS} technique with $\gamma_r$ equal to -10 and -15 dB achieved \ac{INR} levels of -9.686 and -16.116 dB, respectively.

Although the protection criterion was not met for any of the tested radius values, for the 2,400-meter radius, the \ac{INR} level obtained for the 80th percentile was relatively close to the expected value, suggesting that for a slightly larger radius, the reduction in interference achieved would be decreased. However, it is important to note that the extreme \ac{INR} values for the zoning strategy are significantly higher than those obtained for the other two methods, as can be concluded from the \ac{INR} values obtained by the methods at the 90th percentile. In this case, the INR values obtained for all radii of the zoning strategy were 25.94, 22.38, 17.76 and 16.24 dB, starting with the smallest radius and moving to the largest. In contrast, for the \ac{IEIBS} method with a selection threshold $\alpha_e=-10$ dB the obtained INR value is -9.889 dB. For the \ac{IAIBS} method with selection thresholds of $\alpha_r$ equal to -10 and -15 dB the achieved \ac{INR} values were -5.87 and -12.87 dB, respectively. For the \ac{CIEIBS} method with threshold $\gamma_e=-10$ dB the \ac{INR} value was -11.68 dB. For the \ac{CIAIBS} method with limiting threshold $\gamma_r$ equal to -10 and -15 dB the obtained \ac{INR} levels were -8.62 and -14.69 dB, respectively.

In Figure~\ref{fig:sum_of_SEs_for_all_methods}, at the 70th percentile, the sum of \ac{SE} obtained for the baseline scenario is 56.287 bps/Hz. For the same percentile, the sum of \acp{SE} values obtained for \ac{DBZ} techniques were 49.884, 36.871, 10.573 and 5.031 bps/Hz. For the \ac{IEIBS} with $\alpha_e=-10$ dB and the \ac{IAIBS} with $\alpha_r$ equal to -10 and -15 dB the obtained values were 2.926, 31.299 and 20.675 bps/Hz, respectively. For the \ac{CIEIBS} with $\gamma_e=-10$ dB and the \ac{CIAIBS} with $\gamma_r=-10,-15$ dB the obtained values were 2.184, 25.070 and 18.573 bps/Hz.

Figure \ref{fig:Num_admitted_UEs_for_all_methods} shows the number of scheduled \acp{UE} and the occurrence of these numbers for all methods. For the \ac{DBZ} method with radius of 900, 1,500, 2,100 and 2,400 meters the \ac{MNAU} was 8 for both 900 and 1,500 radius, and 4 and 5 for the rest. Those respective values had an occurrence of 33\%, 0.5 \%, 2.5\% and 1.5\%. If the size of the radius is increased, the protection zones get bigger and less \acp{UE} will be admitted in the shared band. For the \ac{IEIBS} technique with selection threshold $\alpha_e=-10$ dB the \ac{MNAU} was 2, with an occurrence of 6.5\%. For the \ac{IAIBS} method with selection threshold $\alpha_r$ equal to -10 and -15 dB the \ac{MNAU} were 7 and 6, respectively, both with an occurrence of 0.5\%. For the \ac{CIEIBS} method with threshold $\gamma_e=-10$ dB, the \ac{MNAU} was 1, with occurrence of 38\%. For the \ac{CIAIBS} method with thresholds $\gamma_r$ equal to -10 and -15 dB the \ac{MNAU} was 6 and 5, respectively, with occurrences of 1.5\% and 6\%.

\section{Conclusion}

The results suggest that the \ac{DBZ} requires very large radius—greater than 2,400 meters—in order to meet the protection criteria at the 80th percentile. However, the \ac{DBZ} method proves to be less efficient than other methods in terms of \ac{INR} levels and \ac{SE} sums. Both \ac{IAIBS} and \ac{CIAIBS} methods with $\alpha_r=-10$ dB and $\gamma_r=-10$ dB did not obtain the required \ac{INR} reduction, but achieve a result close to what was required, while with $\alpha_r=-15$ dB and $\gamma_r=-15$ dB both achieved the required interference reduction. The \ac{IEIBS} and \ac{CIEIBS} techniques with $\alpha_e=-10$ dB and $\gamma_e=-10$ dB also met the protection criterion.

\printbibliography

\end{document}